\documentclass[seceq]{ptptex}

\usepackage{graphicx}

\notypesetlogo                       

\title{
Multiplicity Dependence of Partially Coherent Pion Production\\
in Relativistic Heavy Ion Collisions
}

\author{
Kenji
\textsc{Morita}$^{1,}$\footnote{E-mail:~morita@hep.phys.waseda.ac.jp},%
Shin
\textsc{Muroya}$^{2,}$\footnote{E-mail:~muroya@yukawa.kyoto-u.ac.jp,  Present Address:~Matsumoto University, Matsumoto 390-1295, Japan},
and Hiroki \textsc{Nakamura}$^{1,}$\footnote{E-mail:~naka@hep.phys.waseda.ac.jp}
}

\inst{
$^1$Department of Physics, Waseda University, Tokyo 169-8555,
Japan\\
$^2$Tokuyama University, Shunan, 745-8511,
Japan
}

\recdate{February 4, 2006}

\abst{
 We investigate two- and three-particle intensity correlation functions
 of pions in relativistic heavy ion collisions for different colliding
 energies. Based on three models of particle production, we analyze the
 degree to which the pion sources are chaotic in the SPS S+Pb, Pb+Pb and
 the RHIC Au+Au
 collisions. The ``chaoticity'', $\lambda$, of the two-particle correlation
 functions is corrected for long-lived resonance decays. The effect of the
 partial Coulomb correction is also examined. Although the partially
 coherent model gives a result which is consistent with that of STAR,
 the chaotic fraction does not exhibit clear multiplicity dependence if we
 take into account both the corrected chaoticity and the weight factor of
 the three-pion correlation function.
 The result of the partially-multicoherent model
 indicates an increasing number of coherent sources in higher multiplicity
 events.
}

\begin{document}

\maketitle

\section{Introduction}\label{Sec:intro}
Pion interferometry has been regarded as an indispensable tool in
relativistic heavy ion physics. Two-particle intensity interferometry
can be used to determine the sizes of the collision system. This fact is
known as the Hanbury Brown-Twiss (HBT) effect. For this reason, it has
been used for exploring the space-time
evolution of hot, dense matter created in heavy ion collisions
\cite{Tomasik_qgp3}. In particular, the most recent experiment at the
Relativistic Heavy Ion Collider (RHIC) obtained an interesting result
which is referred to the ``HBT puzzle''
\cite{Heinz_NPA702,Morita_PTP111}. Hydrodynamical models have failed to
reproduce the experimental results of two-particle correlation
functions. Hydrodynamical models are based on the assumption of local
thermal equilibrium. For the early stage of the space-time evolution,
the validity of this assumption is indirectly verified by the
observation of the large
elliptic flow at RHIC \cite{Kolb_PLB500}. However, equilibration in the
final hadronic stage is still ambiguous. Although an exponential particle
spectrum is expected for a thermal source, it does not require a system
that has reached local thermal equlibrium \cite{Rischke_NPA698}.

The two-particle intensity correlation function
$C_2(\boldsymbol{p_1},\boldsymbol{p_2})$ of identical particles provides
information not only
on source sizes but also on the state of the source through the chaoticity,
$\lambda=C_2(\boldsymbol{p},\boldsymbol{p})-1$. The HBT effect does not
exist if the source
is coherent. The chaoticity $\lambda$ is unity for a
perfectly chaotic source and 0 for a
coherent source, which does not induce the HBT effect. In
multi-particle production phenomena of relativistic heavy ion
collisions, the thermalized source can be chaotic and non-thermal
components can be coherent. For example, if disoriented chiral
condensate domains are created, such domains can decay into coherent
pions. Thus the final state pions may carry information regarding how partons
hadronize if the deconfined phase has been created. Therefore, the chaoticity
$\lambda$ is a very important
quantity to investigate the final state in heavy ion collisions. 

However, $\lambda$ cannot be regarded as the true chaoticity in real
(experimental) situations, because many other effects, such as long-lived
resonance decay contributions
\cite{Gyulassy_PLB217,Bolz_PRD47,Heiselberg_PLB,Csorgo_ZPHYS}, Coulomb
repulsions, and particle contaminations \cite{CERES_NPA714}
affect the chaoticity $\lambda$. As an alternative tool, the
three-particle correlation function has been proposed
\cite{Biyajima_PTP84,Heinz_PRC56,Nakamura_PRC60,Nakamura_PRC61}. 
Three-particle
correlations are more useful for this purpose, because long-lived resonances
do not affect the normalized three-pion correlator,
\begin{align}
 r_3&(\boldsymbol{p}_1,\boldsymbol{p}_2,\boldsymbol{p}_3)\nonumber\\
&=\frac{[C_3(\boldsymbol{p}_1,\boldsymbol{p}_2,\boldsymbol{p}_3)-1]
-[C_2(\boldsymbol{p}_1,\boldsymbol{p}_2)-1]
  -[C_2(\boldsymbol{p}_2,\boldsymbol{p}_3)-1]
-[C_2(\boldsymbol{p}_3,\boldsymbol{p}_1)-1]}
  {\sqrt{[C_2(\boldsymbol{p}_1,\boldsymbol{p}_2)-1]
[C_2(\boldsymbol{p}_2,\boldsymbol{p}_3)-1]
[C_2(\boldsymbol{p}_3,\boldsymbol{p}_1)-1]}}, \label{eq:r3-1}
\end{align}
with $C_3(\boldsymbol{p_1},\boldsymbol{p_2},\boldsymbol{p_3})$ being the
three-particle correlation function \cite{Humanic_PRC60}. The
correlation functions are defined as
\begin{equation}
 C_2(\boldsymbol{p_1},\boldsymbol{p_2}) = \frac{W_2(\boldsymbol{p_1},\boldsymbol{p_2})}{W_1(\boldsymbol{p_1})W_1(\boldsymbol{p_2})}
\end{equation}
and
\begin{equation}
 C_3(\boldsymbol{p_1},\boldsymbol{p_2},\boldsymbol{p_3}) =
  \frac{W_3(\boldsymbol{p_1},\boldsymbol{p_2},\boldsymbol{p_3})}
  {W_1(\boldsymbol{p_1})W_1(\boldsymbol{p_2})W_1(\boldsymbol{p_3})}
\end{equation}
with $W_n(\boldsymbol{p_1},\cdots ,\boldsymbol{p_n})$ being the $n$ particle distribution.
The index of the source chaoticity in the three-pion correlator is the weight
factor
\begin{equation}
 \omega=r_3(\boldsymbol{p},\boldsymbol{p},\boldsymbol{p})/2 \label{eq:omega-r3}
\end{equation}
which is unity for a chaotic source. Due to insufficient statistics, the
three-pion correlator measured in experiments to this time is slightly
different from Eq.~\eqref{eq:r3-1},
\begin{equation}
 r_3(Q_3)=\frac{[C_3(Q_3)-1]-[C_2(Q_{12})-1]-[C_2(Q_{23})-1]-[C_2(Q_{31})-1]}
  {\sqrt{[C_2(Q_{12})-1][C_2(Q_{23})-1][C_2(Q_{31})-1]}}, \label{eq:r3}
\end{equation}
where $Q_{ij}=\sqrt{-(p_i-p_j)^2}$ and $Q_3=\sqrt{Q_{12}^2+Q_{23}^2+Q_{31}^2}$.
The weight factor is defined as $\omega=r_3(0)/2$, so that it is
expected to be the same as the definition Eq.~\eqref{eq:omega-r3}. 

As shown in a previous work \cite{Nakamura_PRC66}, more precise
information concerning the source can be extracted using model analyses
combining
two- and three-particle correlations. In this paper, we analyze the
two-pion and three-pion correlations and investigate the chaoticity 
of the pion sources. Firstly, assuming that the main background
contribution is a long-lived
resonance decay and other effects are successfully removed in the
experimental data, we make a correction to account for the long-lived
resonance decays to the chaoticity of the two-pion correlation
functions, with the help of a statistical model. Secondly, we extract the
weight factor $\omega$ through the  simultaneous construction of $C_2$
and $C_3$.
Finally, we carry out model analyses using the ``true'' chaoticity
$\lambda^{\text{true}}$ after applying the resonance correction and the
weight factor
$\omega$. 
In addition to the analysis of Au+Au collisions at the RHIC given in a
previous paper \cite{Morita_3pirhic}, we give analyses on lower
colliding energy collisions at the CERN Super Proton Synchrotron (SPS)
based on the same method. We treat data from 200$A$ GeV (laboratory
system) S+Pb
collisions measured by the NA44 collaboration
\cite{NA44_2pi_PLB302,NA44_2pi_ZPHYS,NA44_3pi_SPb} and data from 158$A$ GeV
(lab. sys) Pb+Pb collisions measured by the NA44 collaboration
\cite{NA44_2pi_PbPb,NA44_3pi_pbpb} and the WA98 collaboration
\cite{WA98_3pi_prl,WA98_pbpb_prc}. On the basis of the results obtained
from these
experimental data, we investigate the multiplicity dependence of the
extracted model parameters. 

 In the next section, we explain the three models used
in this paper. The correction to account for the long-lived resonance
decays is discussed \S3. In \S4, we will give a combined analysis of the
2$\pi$ and 3$\pi$ correlation functions, with the goal of extracting the weight
factor $\omega$. The results of the model analyses are given in \S5.
The paper is summarized in \S6.

\section{Model description}\label{Sec:model}

\begin{figure}[!b]
 \begin{center}
  \includegraphics{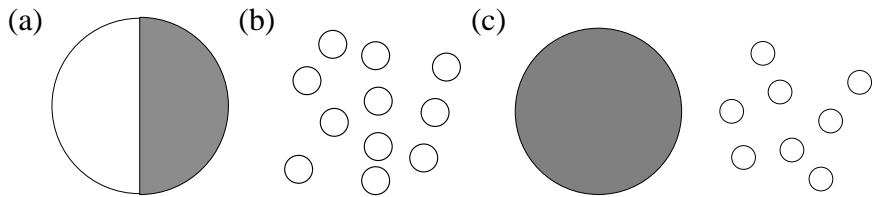}
 \end{center}
 \caption{\label{fig:models}Schematic depiction of Models I--III. The
 shaded area
 denote the chaotic sources. The unshaded area and the small circles
 represent the coherent sources. (a) Model I, (b) Model II, (c) Model
 III.}
\end{figure}

We use three kinds of models in this analysis. In all the models,
$\lambda^{\text{true}}$ and $\omega$ are used as inputs to fix the model
parameters. The output
quantities are the chaotic fraction and the
mean number of coherent sources. Here we do not discuss the origin of
the coherences in these models. Figure \ref{fig:models} presents a
schematic depiction of the models. 

One is a partially coherent model \cite{Heinz_PRC56} whose only
parameter is the chaotic fraction, $\varepsilon_{\text{I}}$, defined as the
ratio of the number of particles coming from the chaotic source to the total
number of particles. In this model, pions are emitted
from a mixture of a chaotic and a coherent source. In general, 
we need to fix the source function in a manner that depends on
the momentum. In the present
analysis, fortunately, the true chaoticity $\lambda^{\text{true}}$ and the
weight factor $\omega$ are given by the correlation functions at
vanishing relative momenta, and hence they can be expressed in terms of
the chaotic fraction.
The relations among
$\varepsilon_{\text{I}}$, $\lambda^{\text{true}}$ and $\omega$ are
\begin{equation}
 \lambda^{\text{true}} = \varepsilon_{\text{I}}(2-\varepsilon_{\text{I}}), 
\quad
 \omega = \sqrt{\varepsilon_{\text{I}}}
 \frac{3-2\varepsilon_{\text{I}}}{(2-\varepsilon_{\text{I}})^{3/2}}.
 \label{eq:modelI}
\end{equation}
We refer to this model as Model I.

The second model, referred to as Model II, is a multicoherent source model
\cite{Nakamura_PRC61}. The parameter in this model is the mean
number of coherent sources, $\alpha_{\text{II}}$, which obeys the Poisson
distribution. Because small coherent sources are independent, a
chaotic source is realized as a cluster of an infinite number of coherent
sources. The parameter $\alpha_{\text{II}}$ is thus related to
$\lambda^{\text{true}}$
and $\omega$ as
\begin{equation}
 \lambda^{\text{true}} = \frac{\alpha_{\text{II}}}{\alpha_{\text{II}}+1},
  \quad
 \omega = \frac{1}{2}\frac{2\alpha_{\text{II}}^2+2\alpha_{\text{II}}+3}
 {\alpha_{\text{II}}^2+3\alpha_{\text{II}}+1}
 \sqrt{\frac{\alpha_{\text{II}}+1}{\alpha_{\text{II}}}} \label{eq:modelII}.
\end{equation}
As shown in Fig.~2 of Ref.~\citen{Nakamura_PRC61}, $\omega$ as a function
of $\alpha_{\text{II}}$ has a minimum value $\omega_{\text{min}}\simeq
0.82$. Hence, there is no
corresponding $\alpha_{\text{II}}$ for $\omega$ smaller than
$\omega_{\text{min}}$.

As each of these models possesses only a single parameter, the
parameters calculated from $\lambda^{\text{true}}$ and $\omega$ should
give the same value if all corrections are correctly made. 

Model III is a ``partially multicoherent'' source model
\cite{Nakamura_PRC61}. This model is a mixture of Model I and Model II,
and its parameters are the chaotic fraction $\varepsilon_{\text{III}}$
and the mean number
of coherent sources $\alpha_{\text{III}}$, which are related to
$\lambda^{\text{true}}$ and $\omega$ as
\begin{align}
 \lambda^{\text{true}} &=
 \frac{\alpha_{\text{III}}}{\alpha_{\text{III}}
 +(1-\varepsilon_{\text{III}})^2},\label{l-pm}
 \\
 \omega &= \frac{2\alpha_{\text{III}}^2 + 2\alpha_{\text{III}}
 (1-\varepsilon_{\text{III}})^2 + 3(1-\varepsilon_{\text{III}})^3
 (1-2\varepsilon_{\text{III}})}
 {2[\alpha_{\text{III}}^2  + 3\alpha_{\text{III}}
 (1-\varepsilon_{\text{III}})^2  +(1-\varepsilon_{\text{III}})^3]} 
 \sqrt{\frac{\alpha_{\text{III}}+(1-\varepsilon_{\text{III}})^2}
 {\alpha_{\text{III}}}}\label{ome-pm}.
\end{align}
In this model, there are two parameters that correspond to two
experimentally measurable quantities, $\lambda^{\text{true}}$ and
$\omega$. Below, we investigate the allowed parameter regions for given sets
of $\lambda^{\text{true}}$ and $\omega$.

\section{Extracting $\lambda^{\text{true}}$ from two-pion correlations}\label{sec:lambda}

In relativistic heavy ion collisions, a non-negligible fraction of pions
comes from the decay of long-lived resonances. Recent analyses based on
statistical models show that hadrons are chemically frozen near the
critical temperature $T_c$ \cite{Braun-Munziger_PLB518}. Short-lived
resonances, such as $\rho$ and $\Delta$, decay before hadrons reach
the kinetic freeze-out or soon after the freeze-out, but long-lived
resonances, such as hyperons, can decay long after the kinetic freeze-out of
pions. In the two-pion correlation function, these long-lived resonances
appear as a spike near $q\sim 0$, whose width is too small to be resolved
with the current experimental resolutions, $\Delta q \simeq 5-10$ MeV. Thus,
chaoticities $\lambda^{\text{exp}}$ measured experimentally 
are smaller than the true chaoticities due to long-lived resonance decays
\cite{Bolz_PRD47,Csorgo_ZPHYS,Wiedemann_PRC56}. Following
Ref.~\citen{Wiedemann_PRC56}, we take into account resonances up to
$\Sigma^*(1385)$. We treat resonances whose widths are less
than 5 MeV as long-lived ones, i.e., 
$K_s^0, \eta, \eta', \phi,\Lambda, \Sigma$ and $\Xi$ are considered
long-lived resonances in the
calculation. Though $\omega$ mesons have an
intermediate width, it is known that they distort the shape of the
correlation function at low $q$ but do not reduce the
chaoticity. \cite{Bolz_PRD47}. 

\begin{table}[t!]
 \caption{\label{tbl:t-mu_SPS}Thermodynamic parameters obtained from the
 particle ratio in nucleus-nucleus collisions at SPS. Here, $h^-$ represents
 the negatively charged hadrons.}
 \begin{center}
  \begin{tabular}{cccccc}\hline
   System &  Ratio & Data [Reference] & $T$ [MeV]& $\mu_{\text{B}}$ [MeV] & $\chi^2/N_{\text{dof}}$\\ \hline
   SPS S+Pb & $p/\pi^+$ & 0.18$\pm$0.03
   \cite{Murray_NA44_NPA566,Gazdzicki_NA35_NPA590} & 173 $\pm$ 2 & 196
   $\pm$ 2 &
   36/7 \\
   & $\bar{p}/p$ & 0.12$\pm$0.02 \cite{Jacak_HDNM} & & & \\ 
   & $\bar{p}/\pi^-$ & 0.024$\pm$0.009
   \cite{Murray_NA44_NPA566,Gazdzicki_NA35_NPA590} & & & \\
   & $(K^-+K^+)/2K_s^0$ & 1.07$\pm$0.03 \cite{DiBari_NPA590}& & & \\
   & $K^+/K^-$ & 1.67$\pm$0.15 \cite{DiBari_NPA590} & & & \\
   & $K_s^0/\Lambda$ & 1.4$\pm$0.1 \cite{DiBari_NPA590} & & & \\
   & $K_s^0/\bar{\Lambda}$ & 6.4$\pm$0.4 \cite{DiBari_NPA590}& & & \\ 
   & $\bar{\Lambda}/\Lambda$ & 0.20$\pm$0.01 \cite{Abatzis_NPA566}& & & \\
   & $\Xi^+/\bar{\Lambda}$& 0.21$\pm$0.02 \cite{Abatzis_NPA566}& & & \\

   \hline
   SPS Pb+Pb & $\bar{p}/p$ & 0.085$\pm$0.009 \cite{Jones_NPA610} & 161
   $\pm$ 4 & 223 $\pm$ 7 & 44/9\\
   & $K_s^0/\pi^-$ & 0.125$\pm$0.019 \cite{Rohrich_Hirschegg97}& & & \\
   & $K_s^0/h^-$ & 0.123$\pm$0.020 \cite{WA97_PLB499}& & & \\
   & $\Lambda/h^-$ & 0.077$\pm$0.011 \cite{WA97_PLB499}& & & \\
   & $\Lambda/K_s^0$ & 0.63$\pm$0.08 \cite{WA97_PLB499}& & & \\
   & $\bar{\Lambda}/\Lambda$ & 0.131$\pm$0.017 \cite{WA97_PLB499}& & & \\
   & $\Xi^-/\Lambda$ & 0.110$\pm$0.010 \cite{WA97_PLB499}& & & \\
   & $\Xi/\bar{\Lambda}$ & 0.188$\pm$0.039 \cite{NA49_JPhys25}& & & \\
   & $(\Xi+\bar{\Xi})/(\Lambda+\bar{\Lambda})$ & 0.13$\pm$0.03
   \cite{NA49_JPhys25_189}& & & \\
   & $\Xi^+/\Xi^-$ & 0.232$\pm$0.033 \cite{NA49_JPhys25}& & & \\
   & $K^+/K^-$ & 1.85$\pm$0.09 \cite{Kaneta_NPA638}& & & \\\hline
  \end{tabular}
 \end{center}
\end{table}

For a chaotic source, the reduced chaoticity $\lambda^{\text{eff}}$ is
given in terms of the
ratio of the number of pions from the long-lived resonances to the total
number of pions as
\begin{equation}
 \sqrt{\lambda^{\text{eff}}} = 1 - \frac{N_\pi^{\text{r}}}{N_\pi},
  \label{eq:lambdaeff}
\end{equation}
where $N_\pi$ is the total number of emitted pions, and $N_\pi^{\text{r}}$ is
the number of pions from the decay of long-lived resonances
\cite{Csorgo_ZPHYS}.
We calculate this ratio with the help of the statistical model. For
midrapidity, the particle ratio can be written as the ratio of the number
densities, i.e.,
\begin{equation}
 \frac{N_i}{N_j} = \frac{n_i^0}{n_j^0}, \label{eq:ratio}
\end{equation}
where 
\begin{equation}
 n_i^0 = \frac{g_i}{2\pi^2}\int_{0}^{\infty}dp \, p^2 
  f(E,T,\mu_{\text{B}},\mu_{\text{S}},\mu_{\text{I$_3$}}),
  \label{eq:density}
\end{equation}
with $f(E,T,\mu_j,\cdots)$ being the equilibrium distribution function and
$g_i$ being the number of the degree of freedom of particle $i$
\cite{Cleymans_PRC60}.
For simplicity, we fix the chemical potential of the third component
of the isospin as $\mu_{I_3} = 0$. Then, the thermodynamic
parameters to be determined are the temperature $T$ and the baryon number
chemical potential $\mu_{\text{B}}$, because the strangeness chemical potential
$\mu_{S}$ is determined from the strangeness neutrality
condition. Results for several collision systems obtained from the
$\chi^2$ fit are shown
in Tables \ref{tbl:t-mu_SPS} and \ref{tbl:t-mu_RHIC}.\footnote{
	 Though experimental data can contain contributions from
	 coherent sources, we assume that particle ratios are not
	 affected by the existence of coherent sources.}

\begin{table}[t!]
 \caption{\label{tbl:t-mu_RHIC}Thermodynamic parameters obtained from
 the particle ratio in Au+Au collisions at RHIC.}
 \begin{center}
  \begin{tabular}{cccccc}\hline
   System &  Ratio & Data & $T$ [MeV]& $\mu_{\text{B}}$ [MeV] & $\chi^2/N_{\text{dof}}$\\ \hline
   RHIC Au+Au & $\bar{p}/p$ & 0.71$\pm$0.05
   \cite{STAR_PbarPratio130}& 158 $\pm$ 9 & 36 $\pm$
   6 & 2.4/5\\
   & $\bar{p}/{\pi^-}$ & 0.072$\pm$0.014 \cite{STAR_pbar,STAR_PLB567}& &
   & \\
   & $K^-/\pi^-$ & 0.146$\pm$0.02 \cite{STAR_PLB595}& & & \\
   & $K^{*0}/h^-$ & 0.042$\pm$0.014 \cite{STAR_PRC66}& & & \\
   & $\bar{K}^{*0}/K^{*0}$ & 0.92$\pm$0.14 \cite{STAR_PRC66}& & & \\
   & $\bar{\Lambda}/\Lambda$ & 0.71$\pm$0.05 \cite{STAR_PLB567}& & & \\
   & $\bar{\Xi}/\Xi$ & 0.83$\pm$0.09 \cite{STAR_PRC66}& & & \\\hline
  \end{tabular}
 \end{center}
\end{table}

From Tables \ref{tbl:t-mu_SPS} and \ref{tbl:t-mu_RHIC}, we find
that the quality of the fit becomes better as the collsion
energy increases. It is reasonable to assume that the
system has reached thermal equilibrium and that the pion source becomes
chaotic as the collision energy increases. Below we see
whether this naive assumption is valid.

Since this particle ratio is obtained from the particle numbers
integrated with the particle momenta, $\lambda^{\text{eff}}$ is
calculated from Eqs.~\eqref{eq:lambdaeff}--\eqref{eq:density} using the
integrated particle numbers.
The measured chaoticity $\lambda^{\text{exp}}$, however, depends on
the average momentum of pion pairs. For example, STAR has measured
$\lambda^{\text{exp}}$ for three bins of the transverse momentum ($k_t$)
\cite{STAR_PRL87}. In this paper, we assume, for simplicity, that
the true chaoticity $\lambda^{\text{true}}$ does not depend on the particle
momentum.\footnote{Note that this is also assumed in Model I--III.} 
Therefore, we need to average $\lambda^{\text{exp}}$ over the momenta in order
to evaluate $\lambda^{\text{true}}$. Assuming the $k_t$ dependence of
the measured
$\lambda^{\text{exp}}$ is dominated by long-lived resonances
[i.e., Eq.~\eqref{eq:lambdaeff}], we obtain the averaged
chaoticity $\overline{\lambda}^{\text{exp}}$ as
\begin{equation}
  \overline{\lambda}^{\text{exp}} 
   = \frac{\displaystyle \sum_{i=1}^{n} \lambda^{\text{exp}}_i
  \int_{i\text{-th bin}} k_t dk_t \left(\frac{dN}{k_t dk_t}\right)^2 }
  {\displaystyle \sum_{i=1}^{n} \int_{i\text{-th bin}} 
  k_t dk_t \left(\frac{dN}{k_t dk_t}\right)^2},
  \label{eq:lambdaav}
\end{equation}
with $\lambda_i^{\text{exp}}$ being the measured chaoticity in the $i$-th
$k_t$ bin. [See the appendix of Ref.~\citen{Morita_3pirhic} for the derivation
of Eq.~\eqref{eq:lambdaav}.] The transverse momentum distribution $dN/k_t
dk_t$, in the above equation is taken from experimental results. For S+Pb
collisions, in
which three-particle correlation data are available only for minimum-bias
data, we calculate $\overline{\lambda}^{\text{exp}}$ by simply averaging
the 
$\lambda^{\text{exp}}$ with different multiplicities. This should not affect
our conclusion, because the data exhibit little multiplicity dependence
\cite{NA44_2pi_ZPHYS}.
For $\lambda^{\text{exp}}$, it is known
that the value of $C_2(\boldsymbol{q})-1$ at $\boldsymbol{q}=0$ depends on the
dimension of the projection onto
$\boldsymbol{q}$-space. In general, the value of $\lambda^{\text{exp}}$
obtained from the
1-dimensional Gaussian fitting differs from that obtained from the
3-dimensional Gaussian fitting because of the projection,
experimental resolution, and other effects, though these should be the
same for an ideal measurement. In this paper, we use the value extracted
from the 3-dimensional Gaussian fitting,
\begin{equation}
 C_2^{\text{fit}}(\boldsymbol{q}) = 1 + \lambda^{\text{exp}}
  \exp\left( -R_{\text{side}}^2 q_{\text{side}}^2
      -R_{\text{out}}^2 q_{\text{out}}^2
      -R_{\text{long}}^2 q_{\text{long}}^2 \right).
  \label{eq:c2fit}
\end{equation}
as $\lambda^{\text{exp}}$.

The true chaoticity
is then given by $\lambda^{\text{true}} =
\overline{\lambda}^{\text{exp}}/\lambda^{\text{eff}}$. 
Results for $\lambda^{\text{true}}$ obtained from various systems are
summarized in Table~\ref{tbl:lambdatrue}. The error on
$\lambda^{\text{true}}$ is the sum of the experimental one on
$\overline{\lambda}^{\text{exp}}$, calculated from the errors on
$\lambda_i^{\text{exp}}$ and $dN/k_tdk_t$, and the errors
propagated from the fit of the thermodynamic quatities at the 1-$\sigma$ level,
shown in Tables \ref{tbl:t-mu_SPS} and \ref{tbl:t-mu_RHIC}.

\begin{table}[t!]
 \caption{\label{tbl:lambdatrue}Summary of $\lambda^{\text{true}}$}
 \begin{center}
  \begin{tabular}{ccccc}\hline
    System & Experiment [Reference] & $\overline{\lambda}^{\text{exp}}$
   & $\lambda^{\text{true}}$ & $\lambda^{\text{true}}_{\text{pc}}$\\ \hline
   SPS S+Pb & NA44, min.bias \cite{NA44_2pi_ZPHYS} & 0.585 $\pm$ 0.06
   &  0.94 $\pm$ 0.06 & 0.7$\lambda^{\text{true}}$\\
   SPS Pb+Pb & NA44 Central \cite{NA44_2pi_PbPb} & 0.55 $\pm$ 0.03 &
   0.98 $\pm$ 0.03 & 0.8$\lambda^{\text{true}}$\\
   SPS Pb+Pb & WA98 Central \cite{WA98_pbpb_prc} & 0.58 $\pm$ 0.04 & 
   1.03 $\pm$ 0.04 & 0.8$\lambda^{\text{true}}$\\
   RHIC Au+Au & STAR Central \cite{STAR_PRL87} & 0.57 $\pm$ 0.06 &
   0.93 $\pm$ 0.08 & 0.8$\lambda^{\text{true}}$ \\\hline
  \end{tabular}
 \end{center}
\end{table}

From Table \ref{tbl:lambdatrue}, we see that the value of
$\lambda^{\text{true}}$
are not so different and become close to unity in all systems. However, it
should be noted that there may be an overestimation in $\lambda^{\text{exp}}$
due to an overcorrection for the Coulomb interaction. 
Because the two-pion correlation function is
affected by the Coulomb interaction between the two detected pions,
there are many issues concerning how the Coulomb interaction can be subtracted
from the observed correlation function
\cite{Pratt_PRD33,Bowler_PLB270,Biyajima_PLB353,Sinyukov_PLB432}.
Recently, it has been shown that
a new procedure called the \textit{partial} Coulomb correction, leads to
significant correction to the source sizes
\cite{PHENIX_HBT200PRL,CERES_NPA714}. Such corrections are known to
partially resolve the ``HBT puzzle'', that $R_{\text{out}}$ becomes smaller
than $R_{\text{side}}$, though this correction does not completely resolve this
puzzle. In this paper, we stress that the partial Coulomb
correction affects not only the source size but also the observed chaoticity,
$\lambda^{\text{exp}}$. For example, CERES reports that the $k_t$ dependent
correction to $\lambda^{\text{exp}}$ reaches 15--40 \%
\cite{CERES_NPA714}. This is a
significant correction. Unfortunately, partially corrected data for the
collisions that we treat in this paper (S+Pb, Pb+Pb at the SPS and Au+Au at
130$A$ GeV)
are not available. Ideally, these data should be treated within a
theoretical approach with a Coulomb correction \cite{Biyajima_PLB601},
but this is beyond the scope of this paper. 
In the further analyses given in \S\ref{sec:results},
we also present the results obtained using ``partially corrected''
$\lambda^{\text{exp}}$ in order to see how the results change when
$\lambda^{\text{exp}}$ is reduced. Following the report from CERES
\cite{CERES_NPA714}, according to which the correction is larger at smaller
$k_t$, and 
considering that the two-particle data used here are those of the lowest
momentum bin, we simply multiplied a correction factor of 0.8
except in the S+Pb case, where the correction factor is set to 0.7, because
only S+Pb data are corrected by the Gamow factor. We denote this
``partially Coulomb corrected''$\lambda^{\text{true}}$ by
$\lambda^{\text{true}}_{\text{pc}}$.

\section{Extraction of the weight factor $\omega$}\label{sec:omega}

\subsection{Procedure}
In the previous section, we extracted $\lambda^{\text{true}}$ from the
two-particle correlation data. Next, we must determine the weight factor
$\omega$ in order to analyze the degree to which the pion sources are
chaotic by using the
models, especially Model III [Eq.~\eqref{l-pm}]. To obtain
$\omega=r_3(0)/2$, we must extrapolate $C_2$ and $C_3$ to
$\boldsymbol{q}=0$. The method of extrapolation is described in a
previous paper \cite{Morita_3pirhic} in detail. Here we briefly review
the procedure.

In experimental papers, $\omega$ is extracted by
simply averaging $r_3(Q_3)$ (NA44 \cite{NA44_3pi_SPb, NA44_3pi_pbpb}, WA98
\cite{WA98_pbpb_prc}), in which there is little dependence on $Q_3$, or
using quadratic and quartic fits (STAR \cite{STAR_3pi}). 
A shortcoming of these methods is that the $Q_3$ dependence of $r_3$ may be
more complex than quadratic or quartic \cite{Nakamura_PRC60}. Such
$Q_3$ dependences result from both an asymmetry of the source and
coherence.\footnote{In Ref.~\citen{Heinz_PRC70}, it is pointed out that
there is a possibility that the standard projection method adopted here
may lead to an artificial momentum dependence of the projected
correlators.} 
In this paper, we assume observed $Q_3$ dependences are due to
coherence. This is an assumption, but it is plausible, because the
asymmetry of the source causes a $Q_3$ dependence of
$r_3$ that is somewhat different from that in the observed $r_3$ data
\cite{Nakamura_PRC60}. We reproduce
$C_2(Q_{\text{inv}})$ and $C_3(Q_3)$ using a common source function with
a set of parameters which is determined by minimizing $\chi^2$ with
respect to the experimental data. Then, we evaluate $C_2(0)$ and
$C_3(0)$. The quantity $r_3(Q_3)$ is also calculated for a
consistency check. For simplicity, we use a spherically symmetric
Fourier-transformed source
function with simulateneous emission, 
$F_{ij} = f_{ij}(|\boldsymbol{q}_{ij}|) e^{i(E_i-E_j)t_0}$, 
in which the
exponential term corresponds to emission at a constant time $t_0$.
The assumption of simulateneous emission should be a good
approximation, because the experimental data suggest emissions of short 
duration throughout the broad range of colliding energies \cite{STAR_PRL87}.
Since the finite emission time duration is not related to
$\lambda^{\text{true}}$ but, rather, to the width of the outward correlation
functions, it does not affect our results below.
For the spatial part, $f_{ij}(|\boldsymbol{q}_{ij}|)$, we try the
following three
kinds of source functions:

\begin{align}
 f_{1,ij}(|\boldsymbol{q}_{ij}|) &= 
 \exp\left(-R^2 |\boldsymbol{q}_{ij}|^2 /2 \right),\label{eq:fgauss} \\
 f_{2,ij}(|\boldsymbol{q}_{ij}|) &= 
 \exp\left(-R|\boldsymbol{q}_{ij}|/2 \right), \label{eq:fexp}\\
 f_{3,ij}(|\boldsymbol{q}_{ij}|) &= 
 \frac{1}{\sqrt{\cosh(R|\boldsymbol{q}_{ij}|)}}\label{eq:fcosh}.
\end{align}
Here, $\boldsymbol{q}_{ij} = \boldsymbol{p}_i - \boldsymbol{p}_j$
and $R$ is a size parameter which is to be determined by
the $\chi^2$ fitting.
The third function, Eq.~\eqref{eq:fcosh}, is chosen so as to be quadratic at
small $|\boldsymbol{q}|$ and exponential at large
$|\boldsymbol{q}|$. 

The two- and three-particle correlation functions are then calculated
using the relations

\begin{align}
 C_2(\boldsymbol{p_1,p_2})&=1+\lambda_{\text{inv}}\frac{f_{12}^2}{f_{11}f_{22}} 
  \label{eq:c2c},\\
 C_3(\boldsymbol{p_1,p_2,p_3}) &= 
 1+\nu\left(\sum_{(i,j)}\frac{f_{ij}^2}{f_{ii}f_{jj}}
 +2\nu_3 \frac{f_{12}f_{23}f_{31}}{f_{11}f_{22}f_{33}}\right) \label{eq:c3c}.
\end{align}

where $\lambda_{\text{inv}}$, $\nu$ and $\nu_3$ are phenomenological adjustable
parameters accounting for the non-trivial coherence effect. 
The summation $\Sigma_{(i,j)}$ here runs over $(i,j)=(1,2),(2,3),(3,1)$.
The quantities $\lambda_{\text{inv}}$, $\nu$ and $\nu_3$ are unity in
the case of a
fully chaotic source. We can set $\nu_3=1$ for a
description of $C_3(Q_3)$ at small $Q_3$ \cite{Morita_3pirhic}.
Then, $\lambda_{\text{inv}}$ and $\nu$ are determined by the
$\chi^2$ fitting, like $R$.
We stress that the $\chi^2$ fit is carried out \textit{simultaneously} for the
two- and three-particle correlation data.  

The results of the $\chi^2$ fittings to the experimental data and the resultant
$\omega$ are listed in Table \ref{tbl:omegafit}.

\begin{table}[t!]
 \caption{\label{tbl:omegafit}Results of the $\chi^2$ fitting to $C_2$
 and $C_3$}
 \begin{center}
 \begin{tabular}[t]{ccccccc}\hline
 system & $f(|\boldsymbol{q}|)$ & $R$ [fm] &  $\lambda_{\text{inv}}$ &
  $\nu $ &   $\chi^2 / \text{dof}$ & $\omega$\\  \hline

  SPS S+Pb & $f_1$ & 4.85$\pm$0.31 & 0.49$\pm$0.04 &
  0.34$\pm$0.04 & 2.3/7 & 0.33$\pm$0.38 \\

  (NA44) \cite{NA44_2pi_PLB302,NA44_3pi_SPb} & $f_2$ & 7.55$\pm$0.84 &
  0.79$\pm$0.10 & 0.61$\pm$0.09 & 3.9/7 & 0.48$\pm$0.45 \\

  & $f_3$ & 8.99$\pm$0.80 & 0.55$\pm$0.06 & 0.40$\pm$0.05 & 1.9/7 &
  0.40$\pm$0.44 \\ \hline

  SPS Pb+Pb & $f_1$ & 7.37$\pm$0.61 & 0.54$\pm$0.05 &
  0.49$\pm$0.07 & 1.2/6 & 1.06$\pm$0.59 \\

  (NA44) \cite{NA44_2pi_PbPb,NA44_3pi_pbpb}  & $f_2$ & 11.4$\pm$1.5 &
  0.83$\pm$0.11 & 0.85$\pm$0.16 & 2.7/6 & 1.16$\pm$0.69 \\
  
  & $f_3$ & 13.8$\pm$1.5 & 0.61$\pm$0.07 & 0.58$\pm$0.10 & 0.9/6 &
  1.15$\pm$0.67 \\ \hline

  SPS Pb+Pb & $f_1$ & 7.73$\pm$0.10 & 0.36$\pm$0.01 &
  0.29$\pm$0.01 & 113/17 & 0.81$\pm$0.12 \\

 (WA98) \cite{WA98_pbpb_prc} & $f_2$ & 14.2$\pm$0.29 & 0.75$\pm$0.02 &
  0.62$\pm$0.02 & 14/17 & 0.68$\pm$0.12 \\

  & $f_3$ & 15.8$\pm$0.28 & 0.47$\pm$0.01 & 0.39$\pm$0.01 & 17/17 &
  0.78$\pm$0.14 \\ \hline

  RHIC Au+Au & $f_1 $& 7.0$\pm$0.07 &
  0.54$\pm$0.01  &
  0.48$\pm$0.01 & 110/30 &0.958$\pm$0.09 \\

 (STAR) \cite{STAR_PRL87,STAR_3pi} & $f_2$ & 14.4$\pm$0.2  &
  1.18$\pm$0.03 & 1.08$\pm$0.03 & 79.7/30 & 0.736$\pm$0.09 \\

  & $f_3$ & 15.2$\pm$0.2 & 0.71$\pm$0.01 & 0.64$\pm$0.02 & 15.8/30 &
  0.872$\pm$0.097 \\\hline
 \end{tabular}
 \end{center}
\end{table}

\subsection{SPS S+Pb}

\begin{figure}[ht]
 \begin{center}
  \includegraphics[width=\textwidth]{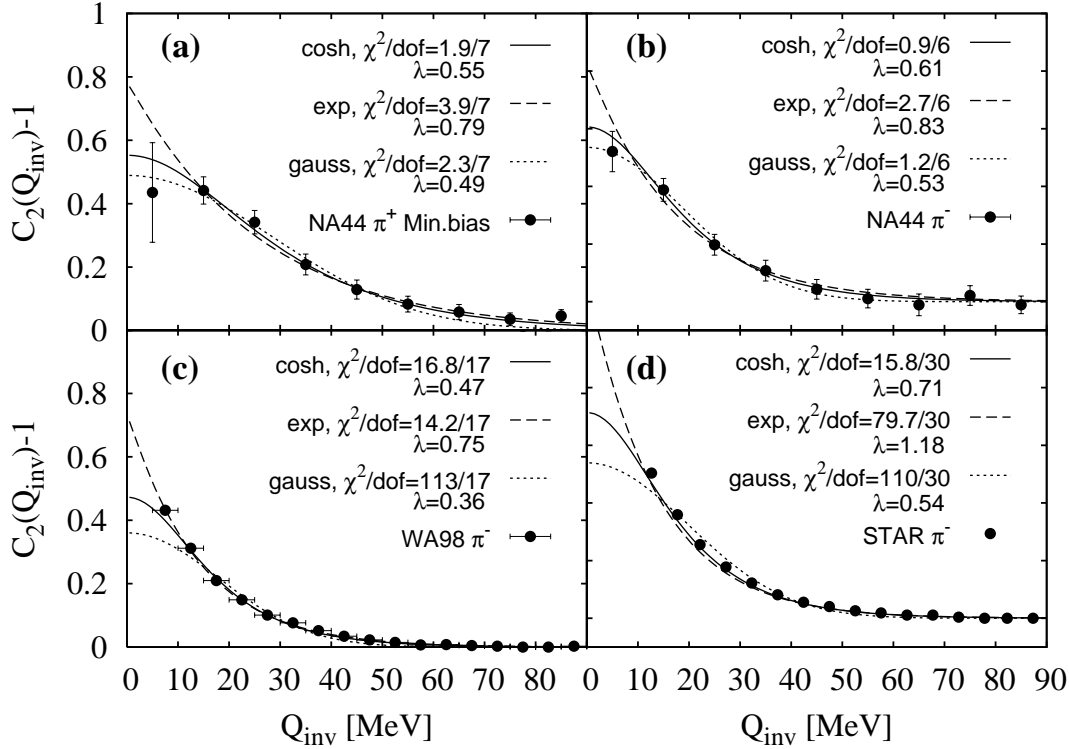}
 \end{center}
 \caption{\label{fig:c2}Two-pion correlation function
 $C_2(Q_{\text{inv}})$ in the (a) S+Pb collisions at the SPS, (b) Pb+Pb
 collisions at the SPS, measured by NA44, (c) Pb+Pb collisions at the
 SPS, measured by WA98, and (d) Au+Au collisions at the RHIC. The lines
 represent our results for the fits of each source function (see
 text). Filled circles
 represent the experimental results (SPS S+Pb by NA44\cite{NA44_2pi_PLB302},
 SPS Pb+Pb by NA44\cite{NA44_2pi_PbPb}and WA98\cite{WA98_pbpb_prc}, and RHIC
 Au+Au by STAR\cite{STAR_PRL87}).}
\end{figure}

\begin{figure}[ht]
 \begin{center}
  \includegraphics[width=\textwidth]{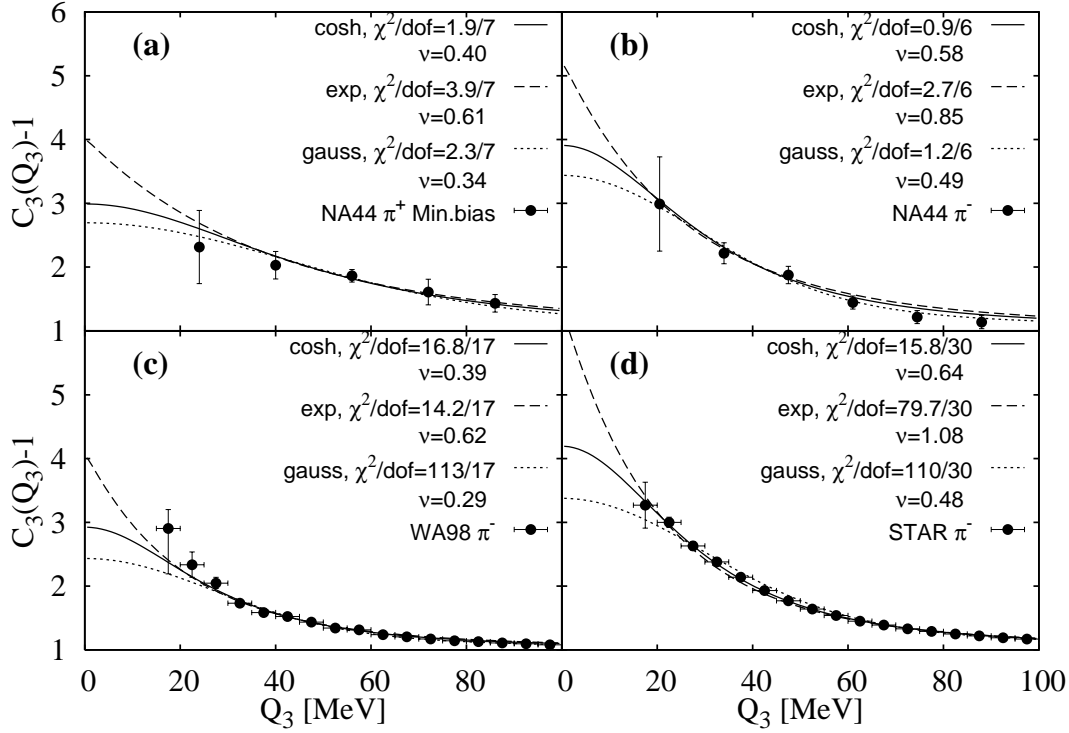}
 \end{center}
 \caption{\label{fig:c3}Three-pion correlation function
 $C_3(Q_3)$ in the various collisions. The identification of the symbols
 is the same as in
 Fig.~\ref{fig:c2}. The experimental results presented in (a)--(d) are
 taken from Refs.~\citen{NA44_3pi_SPb}, \citen{NA44_3pi_pbpb},
 \citen{WA98_pbpb_prc}, and \citen{STAR_3pi}, respectively.}
\end{figure}

\begin{figure}[ht]
 \begin{center}
 \includegraphics[width=\textwidth]{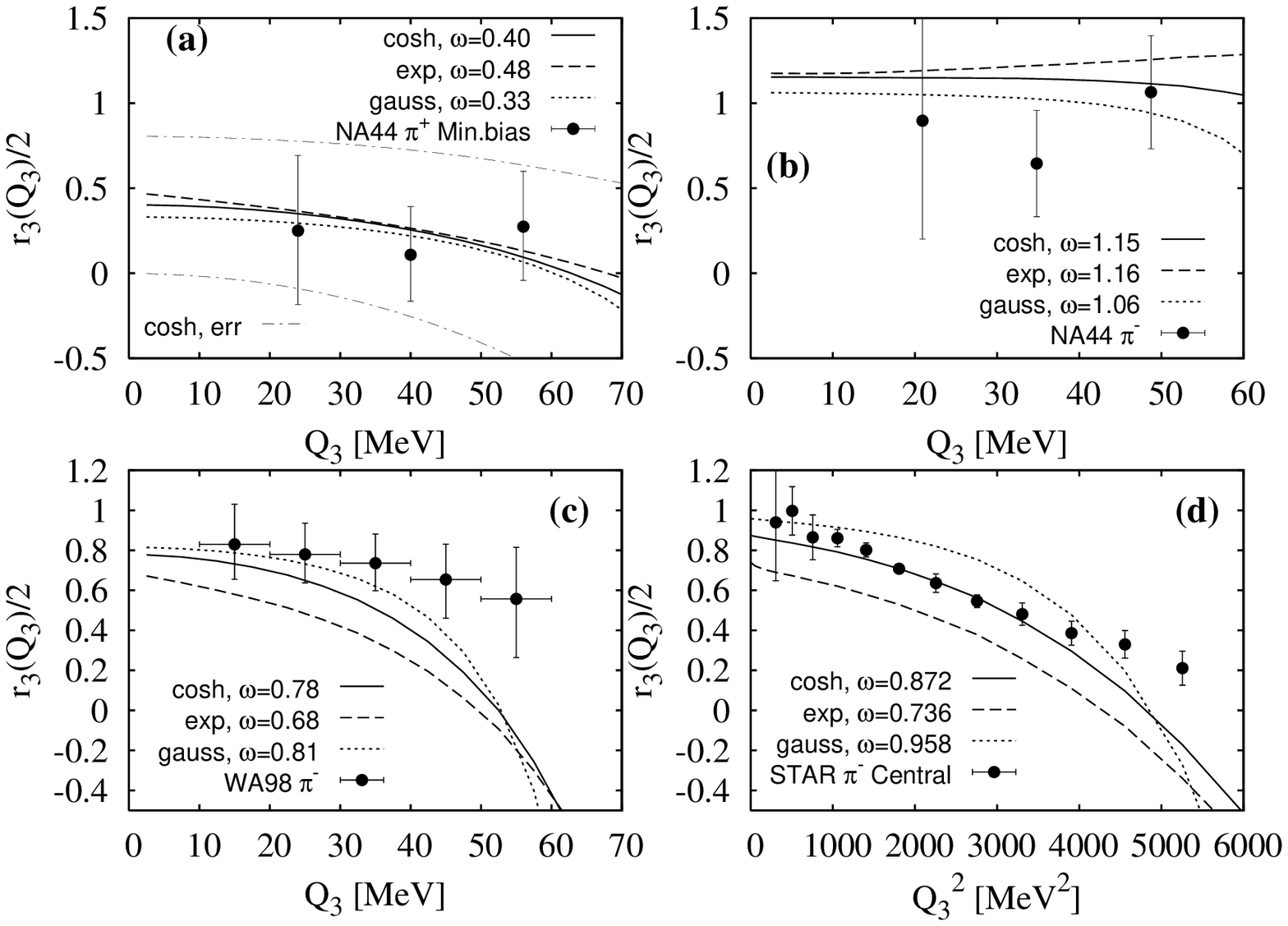}
 \end{center}
 \caption{\label{fig:r3}The normalized three-pion correlator
 $r_3(Q_3)/2$.  The identification of the symbols is the same as in
 Fig.~\ref{fig:c2}. The experimental results presented in (a)--(d) are
 taken from
 Refs.~\citen{NA44_3pi_SPb},\citen{NA44_3pi_pbpb}, \citen{WA98_pbpb_prc},
 and \citen{STAR_3pi}, respectively. The dot-dashed line in (a) represents
 the uncertainty propagated from the fitting parameters for the ``cosh'' case
 (see text).}
\end{figure}

Comparisons of the fitting results and the experimental data are displayed
in Figs.~\ref{fig:c2}(a), \ref{fig:c3}(a) and \ref{fig:r3}(a). The curves in
Figs.~\ref{fig:c2}(a) and \ref{fig:c3}(a) are fitted to the
data using three kind of source functions: Gaussian
[Eq.~\eqref{eq:fgauss}], exponential [Eq.~\eqref{eq:fexp}] and 
cosh [Eq.~\eqref{eq:fcosh}]. The fitting range is
$0 < Q_{ij}, Q_3 < 60$ MeV, which is adjusted to the available $Q_3$
range for $r_3(Q_3)$. In the case of $C_2$, we see
that the experimental data are fit well by a quardratic function at small
$Q_{\text{inv}}$ and an exponential function at large
$Q_{\text{inv}}$. For this reason, the $\chi^2$ values in the
exponential case and the
Gaussian case are larger than those in the cosh case. A similar tendency is
also seen in the case of $C_3$, though the exponential case is not
excluded by the
data point at the smallest $Q_{\text{inv}}$. The value of $r_3$ obtained from
the fitted $C_2$ and $C_3$ do not
differ greatly. With a simple naked-eye extrapolation in
Fig.~\ref{fig:r3}(a), it appears that
$\omega=r_3(0)/2$ becomes $\sim$0.4 for all source functions. Taking the
smallest $\chi^2$ value, we adopt the cosh case and obtain
$\omega=0.40\pm 0.44$. Note that, by the definition of $r_3(Q_3)$, where
$C_2(Q_{ij})-1$ is in the denominator, there exists a large uncertainty in
the calculated $r_3(Q_3)$ value, because of the errors on
$\lambda_{\text{inv}}$. In Fig.~\ref{fig:r3}(a), we plot the error band
for the cosh case. The uncertainty 0.44 associated with $\omega$ is
also obtained by extrapolation of the error band to $Q_3=0$.

\subsection{SPS Pb+Pb}

For Pb+Pb collisions at 158$A$ GeV, we compared our results to data
measured by both NA44 and WA98. Figures
\ref{fig:c2}(b), \ref{fig:c3}(b) and \ref{fig:r3}(b) plot the results
of the $\chi^2$ fit to the NA44 data and
Figs.~\ref{fig:c2}(c), \ref{fig:c3}(c) and \ref{fig:r3}(c) plot those for the
WA98 data.
The fitting ranges are adjusted to the available $r_3(Q_3)$ data range, as
in the S+Pb case, and we use the data for $Q_{ij},Q_3 < 60$ MeV.

Comparing our fit to the NA44 data, we find that all source functions seem to
give nice
descriptions of the data [see Figs.~\ref{fig:c2}(b) and
\ref{fig:c3}(b)]. The statistics are still insufficient, especially
at low $Q_3$, in the three-particle correlation function to discriminate
the best source function. The weight factor $\omega$ is larger than
unity at the best fit value. This is associated with the large
errorbars and consistent with the WA98 case within the errorbars.
For further analysis (given below), we adopt the cosh case because it gives the
best $\chi^2$ value. On the other hand, the WA98 data exhibit better
statistics [see Figs.~\ref{fig:c2}(c) and \ref{fig:c3}(c)]. This
excludes the Gaussian case in the fit to
$C_2$ and $C_3$. For the $\chi^2$ values, the exponential case has the best
value, $\chi^2/\text{dof}=14/17$. This tendency seems to be different
from that in the NA44 case. In Ref.~\citen{WA98_pbpb_prc}, the three-particle
correlation function is fit well by a double-exponential correlation
function. Although the exponential case seems to be the best of the
three, $r_3(Q_3)$ from the exponential source function deviates from the
experimental result [Fig.~\ref{fig:r3}(c)]. 
This should not be regarded as a serious problem, however, because errors on
$\lambda_{\text{inv}}$ and $\nu$ lead to an uncertainty on $r_3(Q_3)$, as
shown in the SPS S+Pb case [Fig.~\ref{fig:r3}(c)]. A likely reason
for this deviation is that $C_3(Q_3)$ is smaller than that obtained from
the experimental data in
the exponential case at low $Q_3$. Naive extrapolation by naked eye in
Fig.~\ref{fig:r3}(c)
again suggests that the cosh or
Gaussian case is better, but this may be a coincidence in the Gaussian case,
because both
$C_2$ and $C_3$ deviate from the experimental data at low relative
momenta. Hence, we adopt the cosh case as the result for $\omega$. 
Despite the different behavior of $C_3(Q_3)$, the result for
$\omega$ obtained from the WA98 data is consistent with the that
obtained from the NA44 data.

\subsection{RHIC Au+Au}

Results presented in this subsection are the same as those presented in
Ref.~\citen{Morita_3pirhic}. 
Figures \ref{fig:c2}(d), \ref{fig:c3}(d) and \ref{fig:r3}(d) display
the correlation functions and the three-pion correlator for the Au+Au
collisions at the RHIC energy. The fitting range is $0 < Q_{\text{inv}} <
90$ MeV for $C_2$ and $0 < Q_3 < 100$ MeV for $C_3$. The high
statistics of the data allows us to discriminate the source function.
In the cosh case, the value of $\chi^2$ is much smaller than other two
cases. Finally, we obtain $\omega=0.872 \pm 0.097$.

\section{Results and discussions}\label{sec:results}

In \S\ref{sec:lambda}, we extracted $\lambda^{\text{true}}$ from 
the experimental data, $\lambda^{\text{exp}}$, with the help of a
statistical model. In \S\ref{sec:omega}, we extracted the weight factor
$\omega$ from the experimental data. Now we have two input
parameters for the analysis employing Models
I--III. [see Eqs.~\eqref{eq:modelI}--\eqref{ome-pm}].
In Models I and II, there is one model parameter
($\varepsilon_{\text{I}}$ for Model I and $\alpha_{\text{II}}$ for Model
II) corresponding to the two input quantities, $\lambda^{\text{true}}$
and $\omega$. We determined the model parameters by minimizing

\begin{equation}
 \chi^2 \equiv
  \frac{[\lambda^{\text{true}}_{\text{exp}}
  -\lambda^{\text{true}}_{\text{cal}}]^2}
  {(\delta\lambda^{\text{true}}_{\text{exp}})^2}
  +
  \frac{[\omega_{\text{exp}}-\omega_{\text{cal}}]^2}
  {(\delta\omega_{\text{exp}})^2},\label{eq:chi2}
\end{equation}
where $\lambda^{\text{true}}_{\text{exp}}$ and $\omega_{\text{exp}}$
were extracted from the experimental data in \S\ref{sec:lambda} and
\ref{sec:omega}. The quantities $\delta\lambda^{\text{true}}_{\text{exp}}$ and
$\delta\omega_{\text{exp}}$ are the errors on
$\lambda^{\text{true}}_{\text{exp}}$ and $\omega_{\text{exp}}$, which are
given in Tables~\ref{tbl:lambdatrue} and \ref{tbl:omegafit}, respectively.
In Models I and II, $\lambda^{\text{true}}_{\text{cal}}$ and
$\omega_{\text{cal}}$ are functions of $\varepsilon_{\text{I}}$ and
$\alpha_{\text{II}}$, respectively, calculated using Eqs.~\eqref{eq:modelI} and
\eqref{eq:modelII}. In Model III, we solve
Eqs.~\eqref{l-pm} and \eqref{ome-pm} for the given sets of
$\lambda^{\text{true}}_{\text{exp}}$ and $\omega_{\text{exp}}$ to obtain
$\varepsilon_{\text{III}}$ and $\alpha_{\text{III}}$. However, solutions
of these equation may exist in unphysical parameter regions, such as
$\varepsilon_{\text{III}} < 0$. In such cases, we determine a ``Best
fit'' solution by minimizing the above $\chi^2$ within the physical
model parameter region, $0 \leq \varepsilon_{\text{III}} \leq 1$ and 
$0 \leq \alpha_{\text{III}} < \infty$.

\subsection{Partially coherent model (Model I)}

\begin{figure}[t!]
 \begin{center}
  \includegraphics[width=3.875in]{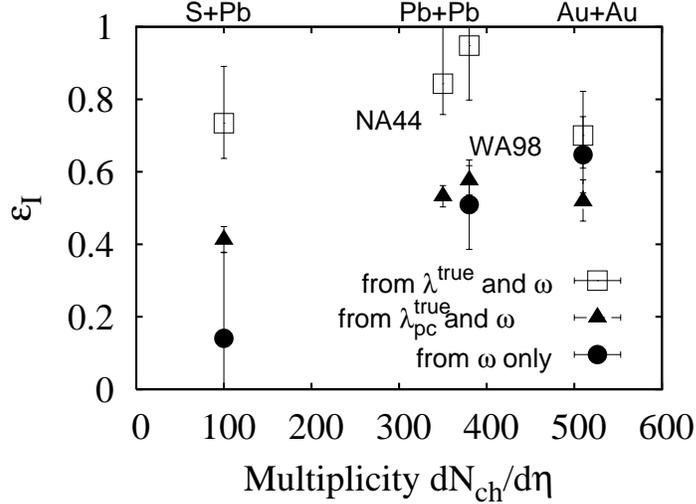}
 \end{center} 
\caption{\label{fig:partial}$\varepsilon_{\text{I}}$
 as a function of the multiplicity. For the Pb+Pb collisions at the SPS,
 we plot both the
 NA44 and WA98 data with a slightly shifted horizontal axis for a clear
 comparison of the results.}
\end{figure}

We plot $\varepsilon_{\text{I}}$ determined from Eq.~\eqref{eq:chi2}
as a function of the multiplicities of various collision systems in
Fig.~\ref{fig:partial}. We consider three cases: The open squares represent the
results obtained from Eq.~\eqref{eq:chi2} using
$(\lambda^{\text{true}},\omega)$, the solid triangles represent those
obtained using
$(\lambda^{\text{true}}_{\text{pc}},\omega)$,  and the closed circles represent
those calculated from $\omega$ only, using the
second equation in Eq.~\eqref{eq:modelI}.

While the result denoted ``$\omega$ only'' exhibits clear increase of
$\varepsilon_{\text{I}}$ with multiplicity, which is consistent with the
result obtained by the STAR\cite{STAR_3pi}, neither
$(\lambda^{\text{true}},\omega)$
nor $(\lambda^{\text{true}}_{\text{pc}},\omega)$ exhibit such a clear
dependence. This is because $\lambda^{\text{true}}$ in
Table~\ref{tbl:lambdatrue} does not display a clear multiplicity dependence
and 
$\delta\lambda_{\text{exp}}^{\text{true}} \ll \delta\omega_{\text{exp}}$. 
The value of $\varepsilon_{\text{I}}$ which minimizes $\chi^2$ is
mostly determined by $\lambda^{\text{true}}$ and
$\lambda^{\text{true}}_{\text{pc}}$ only, i.e., the values of
$\omega$ are not clearly reflected in the minimization of $\chi^2$ due
to the fact that they possess 
larger errors than $\lambda^{\text{true}}$.
If this model is good enough,
and if the experimental background
for the correlation, such as the Coulomb correction is successfully removed,
the result of $(\lambda^{\text{true}}_{\text{pc}},\omega)$ should agree
well with
the ``$\omega$ only'' result. In Fig.~\ref{fig:partial}, these results seem
to be quantitatively consistent. If the experimental
accuracy of the three-particle correlation measures improves, these
results will be more conclusive.

\subsection{Multicoherent model (Model II)}

\begin{table}[ht!]
 \caption{\label{tbl:multic}The parameter $\alpha_{\text{II}}$ in the
 multicoherent model (Model II)}
 \begin{center}
 \begin{tabular}[t]{cccc}\hline
  System & From ($\lambda^{\text{true}},\omega$) & From
  ($\lambda^{\text{true}}_{\text{pc}},\omega$) & From $\omega$ only\\  \hline
  SPS S+Pb & 12.97$^{+57.85}_{-6.16}$ & 1.93$^{+0.41}_{-0.31}$ & --- \\
  SPS Pb+Pb (NA44) & 40.12$^{+\infty}_{-23.84}$ & 3.55$^{+0.67}_{-0.51}$ &
  0.72$^{+0.81}_{-0.38}$ \\
  SPS Pb+Pb (WA98) & 201.93$^{+\infty}_{-179.2}$ & 4.59$^{+1.34}_{-0.92}$
  & --- \\ 
  RHIC Au+Au & 10.54$^{+21.03}_{-4.88}$ &  2.86$^{+1.25}_{-0.76}$ &
  7.59$^{+9.66}_{-6.53}$\\
  \hline
 \end{tabular}
 \end{center}
\end{table}

The results for $\alpha_{\text{II}}$ in the multicoherent model
[Eq.~\eqref{eq:modelII}] are displayed in Table~\ref{tbl:multic}.
As in the case of Model I, three cases are shown. The results for
$(\lambda^{\text{true}},\omega)$ and
$(\lambda^{\text{true}}_{\text{pc}},\omega)$ cases 
mainly reflect their $\lambda^{\text{true}}$ and
$\lambda^{\text{true}}_{\text{pc}}$ values due to the fact that
$\delta\lambda^{\text{true}}$ is much smaller
than $\delta\omega$.
In the $(\lambda^{\text{true}},\omega)$ case, $\alpha_{\text{II}}$ takes
a very large value, coming from $\lambda^{\text{true}}\simeq 1$.
The value of $\alpha_{\text{II}}$ in the
$(\lambda^{\text{true}}_{\text{pc}},\omega)$ case is significantly
smaller, but there is no clear multiplicity
dependence in this case as expected from the fact that
$\lambda^{\text{true}}_{\text{pc}}$ does not possess a clear multiplicity
dependence. In the ``$\omega$ only'' case, there are no solutions for
the SPS S+Pb and Pb+Pb (WA98) data (the blank entries in
Table~\ref{tbl:multic}), because this model has no
corresponding value of $\alpha_{\text{II}}$ below $\omega\simeq 0.82$
[see Fig.~2 of
Ref.~\citen{Nakamura_PRC61}].
This implies that
this model is not suitable for studying multiplicity dependence;
i.e., models with a chaotic background give a better
description of the data.

\subsection{Partially multicoherent model (Model III)}

\begin{figure}[b!]
 \begin{center}
  \includegraphics[width=\textwidth]{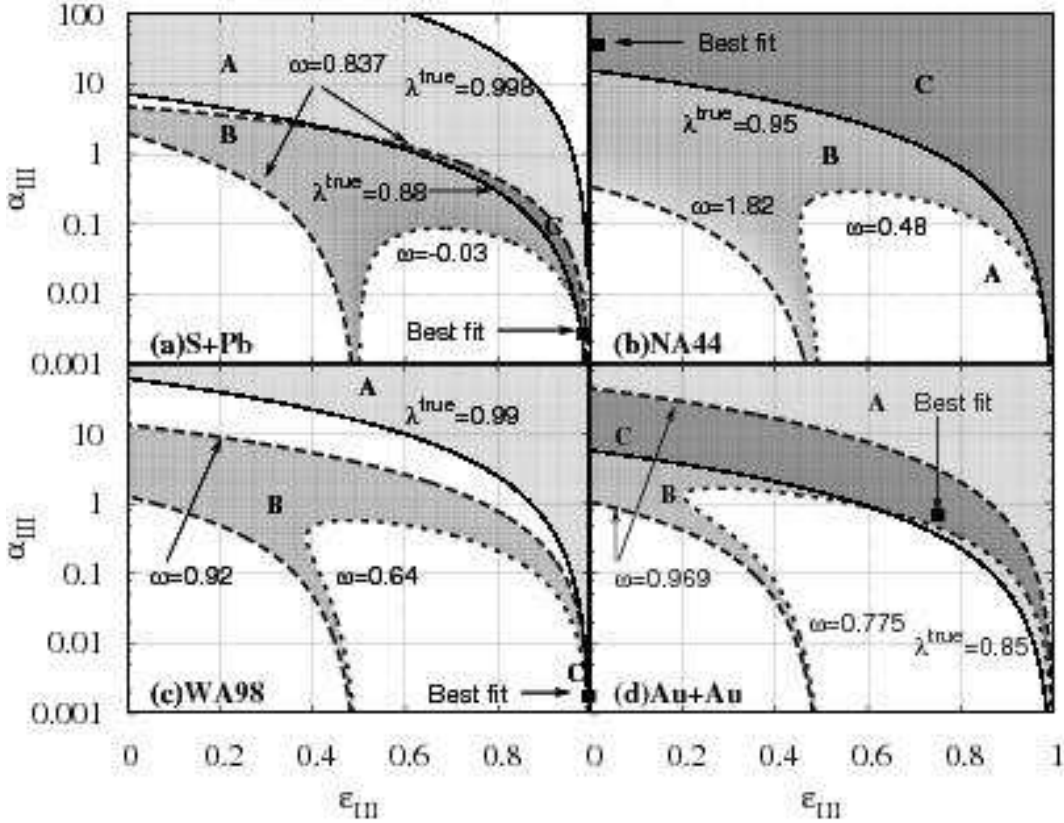}
 \end{center} 
\caption{\label{fig:e-a}Allowed regions for
 $\lambda^{\text{true}}$ in the various systems. (See text for details.)}
\end{figure}

\begin{figure}[ht]
 \begin{center}
  \includegraphics[width=\textwidth]{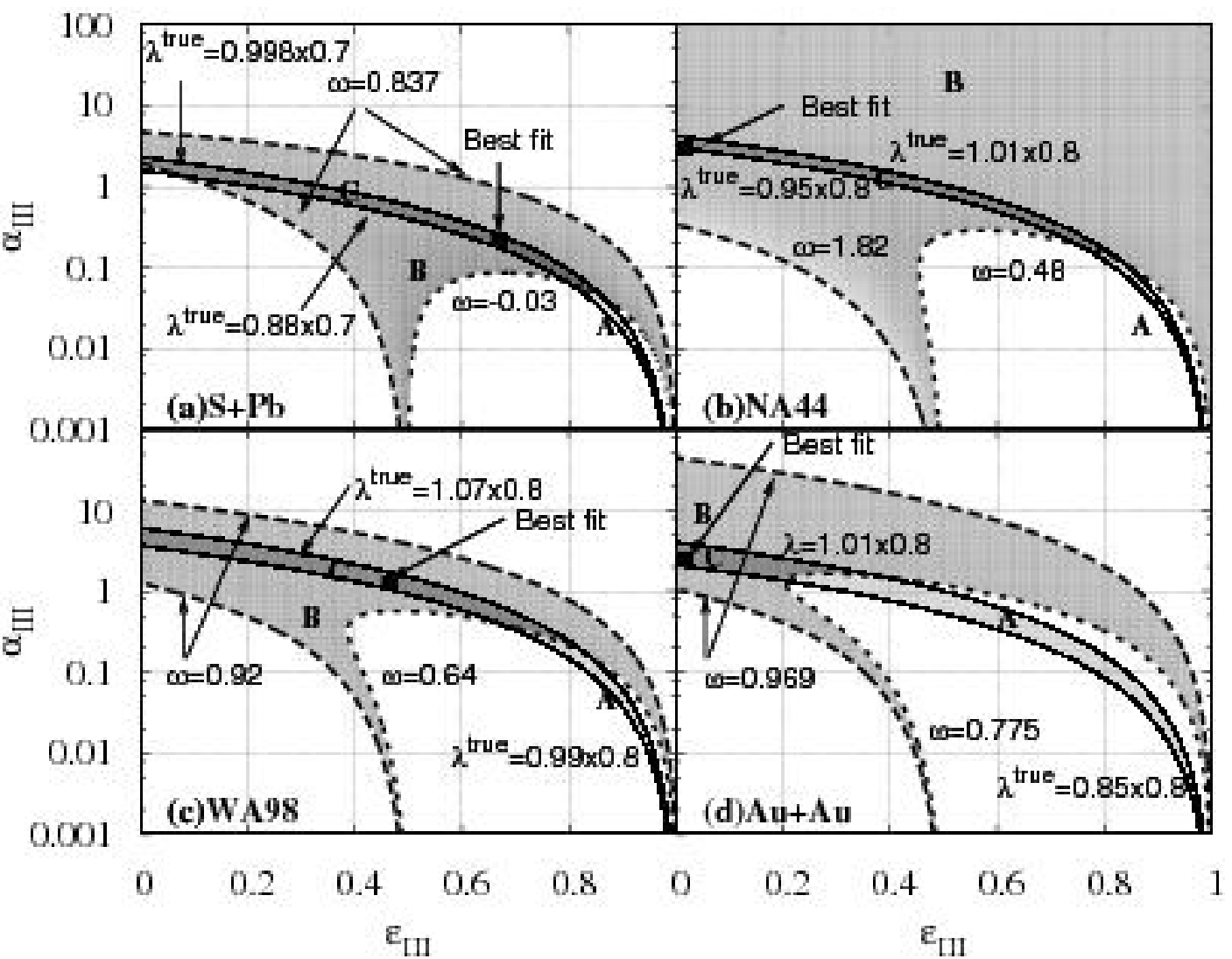}
 \end{center} 
\caption{\label{fig:e-a_lpc}Allowed region for
 $\lambda^{\text{true}}_{\text{pc}}$ in the various systems. (See text
 for details.)}
\end{figure}

In the analysis using Model III, there are two output parameters
($\varepsilon_{\text{III}}$ and $\alpha_{\text{III}}$) corresponding to
the two inputs,
$\lambda^{\text{true}}$ and $\omega$. In the following, we display the
allowed regions of $\varepsilon_{\text{III}}$ and $\alpha_{\text{III}}$
which correspond to the
sets of $\lambda^{\text{true}}$ and $\omega$ in Fig.~\ref{fig:e-a} and
$\lambda^{\text{true}}_{\text{pc}}$ and $\omega$ in Fig.~\ref{fig:e-a_lpc}.

In each figure, the lightly shaded area labeled ``A'' is the allowed
parameter
region corresponding to the value of $\lambda^{\text{true}}$ or
$\lambda^{\text{true}}_{\text{pc}}$, whose
boundary is
indicated by the solid line. Area B, bounded by the dashed (upper
limit of $\omega$) and dotted (lower limit of $\omega$) curves, is the
allowed parameter region corresponding to $\omega$. The darkest area,
which is the region in which Areas A and B overlap, is Area C, which is the
allowed parameter region for $\alpha_{\text{III}}$ and
$\varepsilon_{\text{III}}$. The best fit
values calculated from the values in
Tables \ref{tbl:lambdatrue} and \ref{tbl:omegafit} are indicated by the
squares. 

Figure \ref{fig:e-a}(a) plots the result for the S+Pb collisions using
$\lambda^{\text{true}}$. It is seen that Area C is narrow, and the chaotic
fraction has a lower bound near $\varepsilon_{\text{III}} =0.6$. The best
fit value is $\varepsilon_{\text{III}}\simeq 1$ which reflects that
$\lambda^{\text{true}}$ is near unity and a lower value of $\omega$. If
we adopt a partially Coulomb
corrected
$\lambda^{\text{true}}_{\text{pc}}$, the situation changes. Due to
the smaller
$\lambda^{\text{true}}_{\text{pc}}$[Fig.~\ref{fig:e-a_lpc}(a)],
$\varepsilon_{\text{III}}$ is
allowed for a wider
region and it has a
maximum. The best fit value also shifts to $\varepsilon_{\text{III}} =
0.67$ and
$\alpha_{\text{III}}= 0.21$. Note that, as we can see from
Fig.~\ref{fig:e-a_lpc}, the upper bound of $\varepsilon_{\text{III}}$ is
mainly dominated by the lower bound of
$\omega$.

Figure \ref{fig:e-a}(b) displays the allowed region from
$\lambda^{\text{true}}$ and $\omega$ for the NA44 Pb+Pb collision
dataset. Because the lower bound of $\lambda^\text{true}$ has a large value,
0.95, Area C allows both a mostly chaotic source with a
small number of coherent sources and a number of the coherent sources
with a small chaotic background. In this case, a solution of
Eqs.~\eqref{l-pm}--\eqref{ome-pm} exists in the unphysical region,
$\varepsilon_{\text{III}}=1.01$ and $\alpha_{\text{III}}=0.003$.
Hence, we determine the ``Best fit'' point by minimizing $\chi^2$ in
Eq.~\eqref{eq:chi2}. The result of the minimization gives the ``Best
fit'' at $\varepsilon_{\text{III}}=0$ and $\alpha_{\text{III}}=40.1$.
This result does not always imply that the multicoherent picture is good;
the difference between the value of
$\chi^2$ for this minimum and in another case, for example,
$\epsilon_{\text{III}}=0.99$ and $\alpha=0.01$, is much smaller than
unity. 

If we adopt the partially Coulomb
corrected chaoticity $\lambda^{\text{true}}_{\text{pc}}$, Area C in
Fig.~\ref{fig:e-a_lpc}(b) becomes narrow, as in the S+Pb
case. Because the lower bound of $\omega$ is larger than that in the S+Pb
case, the upper limit of $\varepsilon_{\text{III}}$ becomes smaller.
Similarly, this case does not
have a solution of $\varepsilon_{\text{III}}$ and $\alpha_{\text{III}}$
within the physical region ($\varepsilon_{\text{III}}=1.56$,
$\alpha_{\text{III}}=1.12$), and therefore the ``Best fit'' point is
determined by
minimizing $\chi^2$ in Eq.~\eqref{eq:chi2}. Though the location of the
``Best fit''
point corresponds to the multicoherent picture
($\varepsilon_{\text{III}}=0$), the opposite case
($\varepsilon_{\text{III}}\sim 1$) is
statistically allowed for the same reason as in the previous case.

Similarly, Figs.~\ref{fig:e-a}(c) and \ref{fig:e-a_lpc}(c) display the
results for the WA98 dataset. Since the central value of
$\lambda^{\text{true}}$ exceeds unity, only a very small region near
$\varepsilon_{\text{III}}=1$ is allowed in Fig.~\ref{fig:e-a}(c). This
situation changes drastically if one adopts the
$\lambda^{\text{true}}_{\text{pc}}$.
In Fig.~\ref{fig:e-a_lpc}(c), the allowed region has a shape similar
to that in the NA44 case [Fig.~\ref{fig:e-a_lpc}(b)]. The best fit value is
located at $\varepsilon_{\text{III}} = 0.47$ and $\alpha_{\text{III}}=1.29$. 

Finally, we display the results for Au+Au collisions at RHIC in
Figs.~\ref{fig:e-a}(d) and \ref{fig:e-a_lpc}(d). From
$\lambda^{\text{true}}$ and $\omega$, it is difficult to distinguish the
structure of the source; both a large chaotic fraction with a small number
of coherent sources ($\varepsilon_{\text{III}}\sim 1$ and
$\alpha_{\text{III}}< 1$) and small chaotic fraction with a large number of
coherent sources ($\varepsilon_{\text{III}}\sim 0$ and
$\alpha_{\text{III}}> 1$) can reproduce the experimental data. The best
fit value
is located at $\varepsilon_{\text{III}}=0.75$ and
$\alpha_{\text{III}}=0.77$. However,
if we use the partially Coulomb corrected two-particle correlation
data, the allowed region is strongly restricted. Although the solution
of Eqs.~\eqref{l-pm}--\eqref{ome-pm} exists in the unphysical region,
$\varepsilon_{\text{III}}=-0.54$ and $\alpha_{\text{III}}=6.83$, the
``Best fit'' point is located at $\epsilon_{\text{III}}=0$ and
$\alpha_{\text{III}}=2.86$. The maximum value allowed
for $\varepsilon_{\text{III}}$ is 0.4. This means that the strong chaotic
behavior observed at the RHIC is due to the production of a cluster of coherent
sources.

\begin{figure}[b!]
 \begin{center}
  \includegraphics[width=3.875in]{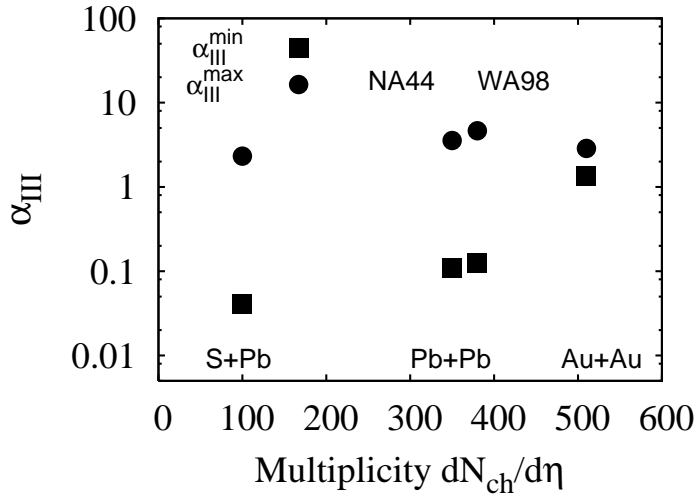}
 \end{center} 
\caption{\label{fig:partialmulti}The minimum (solid squares) and
 maximum (solid circles) value of $\alpha_{\text{III}}$ in Model III
 for various
 collisions as a function of the multiplicity. For the Pb+Pb collisions at
 the SPS, we plot both the NA44 and WA98 data with a slightly shifted
 horizontal axis for a clear comparison of the results.}
\end{figure}

As shown in the
Fig.~\ref{fig:e-a_lpc}, the maximum bounds on
$\varepsilon_{\text{III}}$ are mainly
determined by the lower bounds on $\omega$, which become larger as the
energy and the multiplicity increase. The fact that
$\alpha_{\text{III}}$ becomes larger as the multiplicity increases is
consistent with the result of
a previous analysis by one of the authors (H.~N.), in which the chaoticity
and the weight factor were experimentally
determined.\cite{Nakamura_PRC66} In Fig.~\ref{fig:partialmulti}, we plot
the
maximum and minimum values of $\alpha_{\text{III}}$ for the sets of
$\lambda^{\text{true}}_{\text{pc}}$ and $\omega$ extracted from
Fig.~\ref{fig:e-a_lpc}. We can see that minimum value of
$\alpha_{\text{III}}$
increases with the multiplicity. This suggests that, as the collision energy
and the multiplicity increase, the number of coherent sources
increases. Note that the maximum values are the same as
those obtained using Model II for
$\lambda^{\text{true}}_{\text{pc}}$ (see Table~\ref{tbl:multic}), because
$\alpha_{\text{III}}$ takes its maximum value when
$\varepsilon_{\text{III}}=0$. 
Furthermore, this tendency seems
to be correlated with the plausibility of the statistical model
($\chi^2/N_{\text{dof}}$), given in
Tables~\ref{tbl:t-mu_SPS} and \ref{tbl:t-mu_RHIC}. Though we do not know any
explicit relation between Models
I--III and the hadronization mechanism, this result may reflect a possible
hadronization mechanism from the quark-gluon plasma phase created in the
collisions. 

\section{Summary}
In summary, we have investigated the degree to which the pion sources
are chaotic in
various heavy ion collisions by analyzing the two- and 
three-particle correlation data with three kinds of particle
production models. For the two-particle correlation data, we have
extracted the ``true'' chaoticity by considering long-lived resonance
contributions to the pion multiplicity, with the help of a statistical
model. Using simple source functions, we simultaneously investigated
the two- and three-particle correlation functions to extract the
weight factor $\omega$ of the three-particle correlator. Incorporating the
chaoticity and the weight factor into the models, we have studied
the chaotic fraction and mean number of the coherent sources.
The results for $\varepsilon_{\text{I}}$ obtained from $\omega$
indicates that the system becomes chaotic as the multiplicity
increases. This result is consistent with Ref.~\citen{STAR_3pi}. 
From a multicoherent source point of view, it is concluded that pions at higher
collision energies may be emitted from a cluster of coherent sources
and the number of sources increases as the collision energy and
the multiplicity increase (Fig.~\ref{fig:partialmulti}).

\section*{Acknowledgements}
 The authors would like to acknowledge Professors I.~Ohba and H.~Nakazato for
 their insightful comments.
 This work was partially supported by the
 Ministry of Education, Science and Culture, of Japan (Grant No.13135221
 and Grant No.18540294)
 , Waseda University Grant for Special Research Projects No.~2003A-095
 and 2003A-591, and a Grant for the 21st Century COE Program at Waseda
 University from Ministry of Education, Science and Culture, of Japan.
 One of the author (S.~M.) would like to thank the YITP computer room.

%

\end{document}